\documentclass{iau}
\usepackage{graphicx}
\usepackage{natbib}
\newcommand {\apj} {ApJ}
\newcommand {\apjs} {ApJS}
\newcommand {\apjl} {ApJL}
\newcommand {\aap} {A\&A}

\newcommand {\baas} {BAAS}

\newcommand {\mnras} {MNRAS}

\newcommand {\nat} {Nature}

\newcommand {\araa} {Annu. Rev. Astron. Astrophys.}

\newcommand {\jqsrt} {J. Quant. Spectrosc. Rad. Transfer}

\newcommand {\rmxaa} {Rev. Mex. Astron. Astrophys.}

\newcommand {\icarus} {Icarus}

\newcommand {\ssr} {Space Sci. Rev.}

\title[Introduction] 
{Laboratory astrophysics: key to understanding the Universe}

\author[Ewine F. van Dishoeck]   
{Ewine F. van Dishoeck$^{1,2}$}

\affiliation{$^1$Leiden Observatory, Leiden University, the Netherlands \\ email: {\tt ewine@strw.leidenuniv.nl} \\[\affilskip]
$^2$Max Planck Institute for Extraterrestrial Physics, Garching, Germany}

\pubyear{2019}
\volume{350}
\jname{Laboratory Astrophysics: from Observations to Interpretation} 
\editors{F. Salama, H.J. Fraser, H. Linnartz, ed.}
\begin{document}

\maketitle

\begin{abstract}
  This brief overview stresses the importance of laboratory data and
  theory in analyzing astronomical observations and understanding the
  physical and chemical processes that drive the astrophysical
  phenomena in our Universe.  This includes basic atomic and molecular
  data such as spectroscopy and collisional rate coefficients, but
  also an improved understanding of nuclear, plasma and particle
  physics, as well as reactions and photoprocesses in the gaseous and
  solid state that lead to chemical complexity and building blocks for
  life. Systematic laboratory collision experiments have provided
  detailed insight into the steps that produce pebbles, bricks and
  ultimately planetesimals starting from sub-$\mu$-sized
  grains. Sample return missions and meteoritic studies benefit from
  increasingly sophisticated laboratory machines to analyze materials
  and provide compositional images on nanometer scales. Prioritization
  of future data requirements will be needed to cope with the
  increasing data streams from a diverse range of future astronomical
  facilities within a constrained laboratory astrophysics budget.



\end{abstract}

\firstsection 
\section{Introduction}
\label{sec:intro}

Modern astrophysics is blessed with an increasing amount of high
quality observational data on astronomical sources, ranging from our
own Solar System to the edge of the Universe and from the lowest
temperature clouds to the highest energy cosmic rays. Spectra
containing thousands of features of atoms, molecules, ice and dust are
routinely obtained for stars, planets, comets, the interstellar medium
(ISM) and star-forming regions, and now even for the most distant
galaxies. Realistic models of exo-planetary atmospheres require
information on billions of lines. Theories of jets from young stars
benefit from plasma experiments to benchmark them. Stellar evolution
theories and cosmology rely heavily on accurate rates for nuclear
fusion reactions. The first stars could not have formed without the
simplest chemical reactions producing H$_2$ and HD in primordial
clouds.  Particle physics is at the heart of finding candidates for
the mysterious dark matter.

\begin{figure}[t]
\begin{center}
\includegraphics[width=14cm]{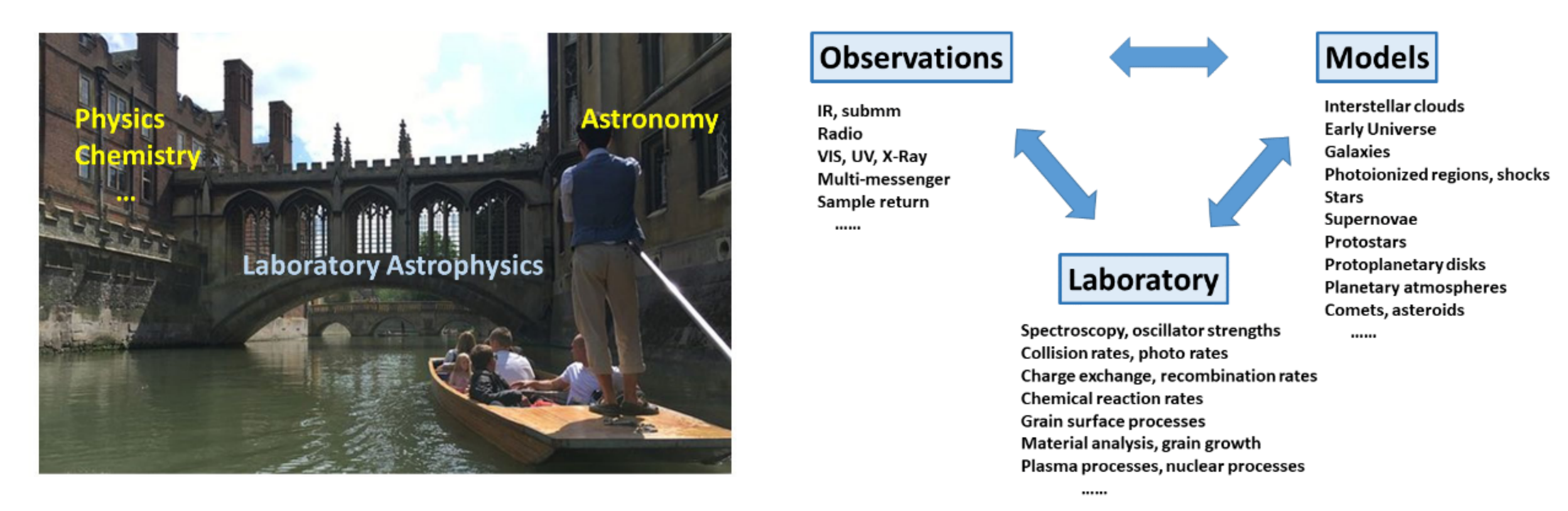} 
 \caption{Left: The Bridge of Sighs in Cambridge (UK), symbolizing
   laboratory astrophysics as a bridge between astronomy and physics,
   chemistry and other sciences. Credit: Tripadvisor.co.uk. Right: the
   triangle of observations, models and laboratory astrophysics, with
   the latter underpinning both.  }
\label{fig:bridge}
\end{center}
\end{figure}

Taken together, there is no doubt that laboratory astrophysics, with
`laboratory' defined to include theoretical calculations, remains at
the foundation of the interpretation of observations and truly `makes
astronomy tick'. It also provides an important bridge between
astronomy and physics, chemistry and other sciences, appropriately
represented in Fig.~\ref{fig:bridge}
with laboratory astrophysics providing the `sigh of relief'. It is
important to recognize that this is not a one-way `bridge', but that
astronomy and physics, chemistry can mutually stimulate and enrich
each other \citep{Dalgarno08}.

This paper will provide some general thoughts about the field, and
then briefly describe a number of recent examples of observational
developments where the availability of new laboratory data has been
important in the analysis.  Often, a comparatively minor investment in
basic studies can greatly enhance the scientific return from expensive
missions.  Examples are included from (i) atomic physics; (ii)
nuclear physics; (iii) molecular physics; (iv) solid-state and
condensed matter physics; and (v) planetary sciences, topics which are
also covered at this Symposium.
Not included here are (vi) plasma physics and (vii) (astro)particle
physics. A comprehensive review of all of these aspects of Laboratory
Astrophyscs is provided by \citet{Savin12}, with some updates in
\citet{Savin19}. The molecular part is well covered in reviews by
\citet{Tielens13,vanDishoeck13,Cuppen17} and the 2013 special issue of
Chemical Reviews, whereas an overview of exoplanetary atmosphere cases
can be found in \citet{Fortney19}. This paper also makes good use of
on-line presentations from recent workshops of the AAS Laboratory
Astrophysics Division. Only limited (and highly incomplete) references
and examples are given throughout this paper.

In 2019, the IAU celebrates its 100 yr existence \footnote{{\tt
    www.iau-100.org}}. The exhibition highlights major discoveries
over the century, many of them enabled by laboratory astrophysics.
This is also the UN International Year of the Periodic Table of
Chemical Elements \footnote{{\tt www.iypt2019.org}}, appropriately reminding
people worldwide that the elements in our body originate from nuclear
reactions in stars.

\section{Laboratory astrophysics as a field}
\label{sec:field}

{\underline {\it Some history.}} Astronomy and laboratory astrophysics have
gone hand in hand since the earliest spectroscopic observations. The
Fraunhofer lines in the spectrum of the Sun, first seen in 1814, could
not be understood without basic atomic spectroscopy. The solar
spectrum also provides an excellent example of the opposite case:
unidentified lines led to the discovery of Helium in a laboratory on
Earth in 1868. Huggins used spectroscopy around 1864 to demonstrate
that there are two different types of nebulae: those that resembled the
Orion nebula which is full of (electric dipole) forbidden lines and
those like the Andromeda galaxy that resemble the spectra of
stars. The [O I] atom with its $^3P$-$^1D$ 6331 \AA \ red (so-called
`nebular) and $^1D$-$^1S$ 5577 \AA \ green (`auroral') lines is prime
example of a system for which deep knowledge of its spectroscopy and
excitation serves both the astronomy and aeronomy communities.

Another early example is provided by the diffuse interstellar bands,
first seen in optical spectra of bright early-type stars by
\citet{Heger22}. Modern instruments have recorded more than 500
different lines, the majority of which are still unidentified after a
century \citep{Cox17}.

{\underline {\it Importance of new facilities.}} Major new facilities
and technology drive the field. On the laboratory side, this includes
large synchrotron and advanced light sources as well as the most
powerful computers, often offered as national facilities across the
world. At the individual institute and researcher level, there are
innovative laboratory set-ups including, for example, cavity ringdown
and CHIRP spectroscopy, He droplets and crossed beam experiments, and
UHV surface science techniques.  On the astronomical side, the field
has been fortunate to have powerful 8-10m optical ground-based
telescopes, UV and X-ray spectroscopy from space with $HST, XMM,
Chandra$ and soon $XRISM$, and (far-)infrared and submillimeter with
{\it Spitzer}, {\it Herschel}, SOFIA and ALMA. With {\it JWST}, ELTs
and many other missions on the horizon providing increasingly sharp
and sensitive data, the need for further laboratory data will surely
increase.

{\underline {\it Funding challenges.}} In spite of the enormous
opportunities, there are some worrying signs. Budgets for laboratory
astrophysics are under stress. Several decades ago, most of the
funding for experiments or theory came from physics or chemistry
because the astronomical questions were driving fundamental new
insights in those fields. However, hot topics in physics and chemistry
have now moved to other areas that have little connection with
astronomy. While these fields occasionally still provide funding for
new equipment or sophisticated computer codes and packages that can
subsequently also be used to address astrophysical questions, the
manpower to carry out these `routine' experiments or calculations has
to come from other sources, most notably astronomy itself.

This then runs into another problem, namely that the current
generation of astronomers is used to large turn-key software packages
and hardly appreciates the importance of the basic data that goes into
them, as well as their uncertainties. Huge amounts of computer time
are spent on MCMC or Bayesian fits to observational data using a
`black box' program to infer physical and chemical parameters such as
abundances, temperature, density, radiation field, filling factor
etc., but the answer is only as good as the basic data that go into
the model.

{\underline {\it Citation pyramid.}} Take as an example the CLOUDY
model and package for photoionized regions \citep{Ferland17}. Many
thousands of manyears and decades of work have gone into computing and
compiling all the atomic data (line positions, oscillator strengths,
collisional and photoionization cross sections, recombination rate
coefficients) that enter the program. Each of these thousands of
papers has a few tens of citations each. The CLOUDY program, widely
used by the community, has some 5000 citations total. In contrast, the
science enabled by CLOUDY has in total well over a million citations.
To give credit where credit is due, astronomers should cite as much as
possible the essential basic data that went into the analysis,
especially since the funding for getting those data is at least
loosely correlated with number of citations in many countries.

{\underline {\it Public codes and databases.}} Codes such as CLOUDY,
SPEX and CHIANTI illustrate another issue, namely that the cost of
writing the initial version is a tiny fraction of the long-term cost
for maintenance, improvement, support and training the next generation
in the use of such a code (Brickhouse, LAW 2018 workshop). Providing a
public code can readily become ones career if the program is widely
used and it is tough to get an academic position for what is regarded
as just a `service' to the community. The latter characterization is
unfair since the code enables frontline science, including by its
author(s), but the label sticks all too easily and drives such
`service providers' to national centers or institutes, or worse, out
of the field so that critical expertise entirely is lost.

Populating the databases with new experiments or calculations requires
a critical evaluation of the numbers and their uncertainties: the
latest value is not always the best! Also, extrapolations are often
needed, for example to lower or higher temperatures. Code comparisons
are very time consuming but should be done on a regular basis and can
be highly valuable in revealing not only discrepancies but also
sensitivities to certain parameters that were not realized
before. Integration of databases and codes with astrophysical tools
requires teams (and referees!) that understand both aspects.

{\underline {\it Organizing the community.}} To have a voice in policy making
and convincing funding agencies to provide more resources where
needed, organization of the community is important. This is happening
now at various national and international levels. In particular, the
IAU has a long-standing tradition in this area, first with the old
Commission 14 on `Atomic and Molecular Data' and after the
restructuring in 2012-2015 with Commission B5 on `Laboratory
Astrophysics'. On the molecular side, it has considerable synergy with
the former Working Group and now Commission H2 on Astrochemistry.

\section{Atomic and nuclear physics}

Atomic physics has long been essential in determining and analyzing
solar and stellar abundances and opacities, supernova remnants,
accretion shocks onto young stars, starburst galaxies and AGN. Extreme
wavelength precision of atomic thorium/argon and iodine lines is now
also key in providing the reference frame for measuring radial
velocities of exoplanets to tens of cm precision.

 \begin{figure}[t]
\begin{center}
\includegraphics[width=11cm]{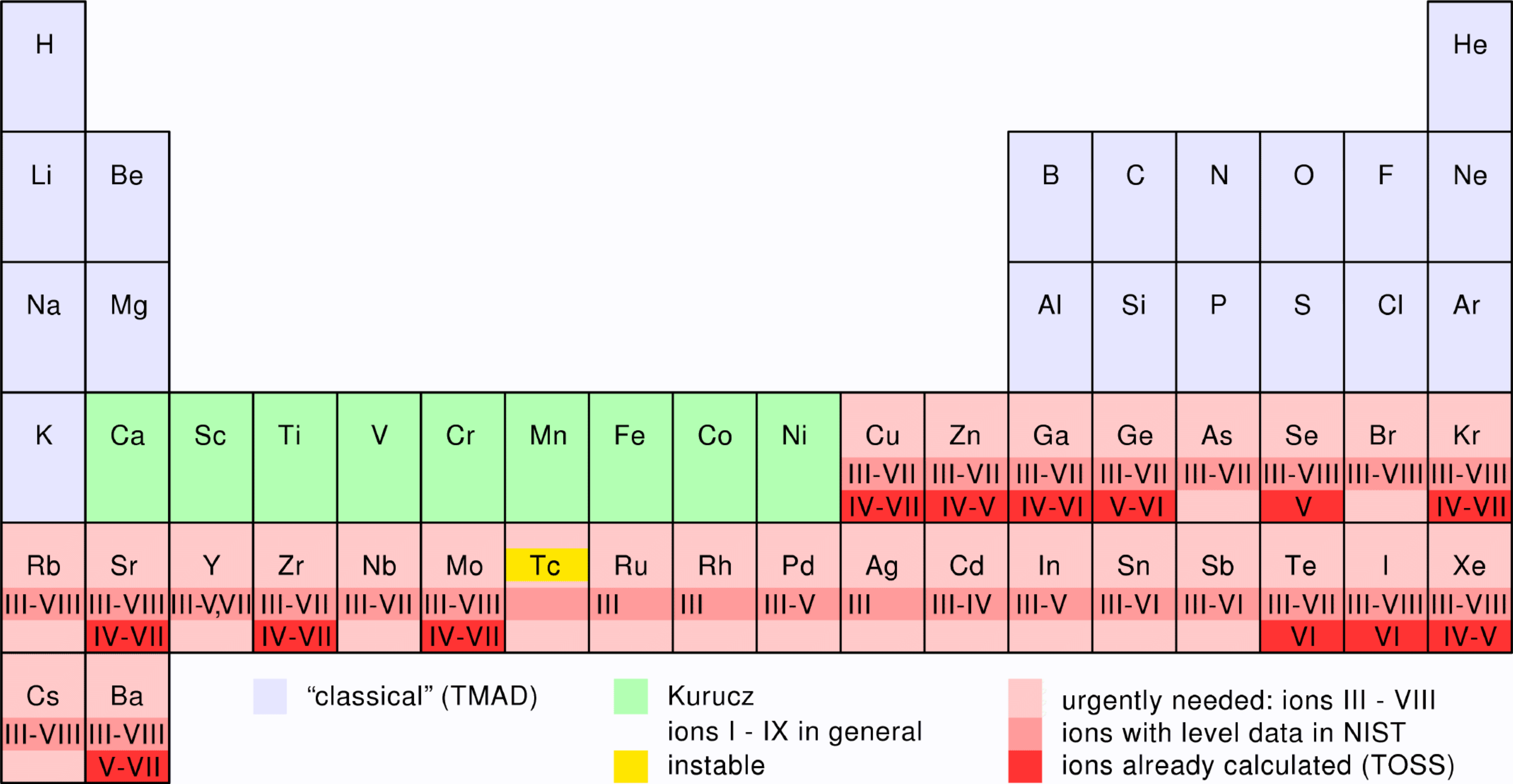} 
  \caption{Summary of elements for which accurate oscillator strengths are still lacking. Credit: T. Rauch.}
\label{fig:elements}
\end{center}
\end{figure}

One recent high energy example is provided by a weak feature around
3.62 keV found in stacked XMM spectra of AGN that was not well fitted with
available spectral synthesis packages \citep{Bulbul14}.  This led to
speculation: was this a sign of new physics, such as sterile neutrino
dark matter?  Or could the feature be explained by previously missed
dielectronic recombination features, e.g., of Ar XVII?  The higher
spectral resolution {\it Hitomi} microcalorimeter spectrum of the
Perseus cluster, however, does not show deviations from (updated)
spectral packages \citep{Aharonian17}.

High-resolution images from {\it Chandra} combined with data from {\it
  NuSTAR} beautifully reveal the production of elements through
nuclear fusion in stars and supernovae ejecta, including, for example,
$^{44}$Ti \citep{Grefenstette17}. An overview of nuclear astrophysics
needs can be found in \citet{Wiescher12}.  Particularly exciting are
the recent data on neutron star mergers, detected through
gravitational waves with optical spectroscopy follow-up, showing that
they are indeed a major source of elements heavier than Fe through the
rapid $r-$process. More generally, surveys of abundances of heavy
elements like Ga and Ge in a wide range of sources (diffuse clouds,
white dwarfs, planetary nebulae) can constrain the relative
contributions of slow (AGB) vs rapid (supernovae, neutron star
mergers) neutron capture processes to current epoch Galactic chemical
evolution \citep[e.g.,][]{Ritchey18}. Such studies are only possible,
however, with accurate values for the oscillator strengths
($f-$values) of the transitions. Fig.~\ref{fig:elements} summarizes
the data needs for heavy elements.

\section{Molecular physics: gas phase}
\label{sec:molgas}

Molecules, both in the gas and in the solid state, are observed
throughout the Universe, from the highest redshift galaxies to the
comets and planets in our own Solar System. Molecular data are needed
to interpret observations of the diffuse interstellar medium, dark
clouds, star-forming regions, protoplanetary disks, exoplanetary
atmospheres, solar system objects, evolved star envelopes, late type
stellar atmospheres, nearby and distant galaxies, and the early
Universe. Besides line frequencies, line strengths, and collisional
data, molecules have the added complexity of a wide range of chemical
reactions that can occur for which rate coefficients are needed over a
large range of conditions.

 \begin{figure}[t]
\begin{center}
\includegraphics[width=12cm]{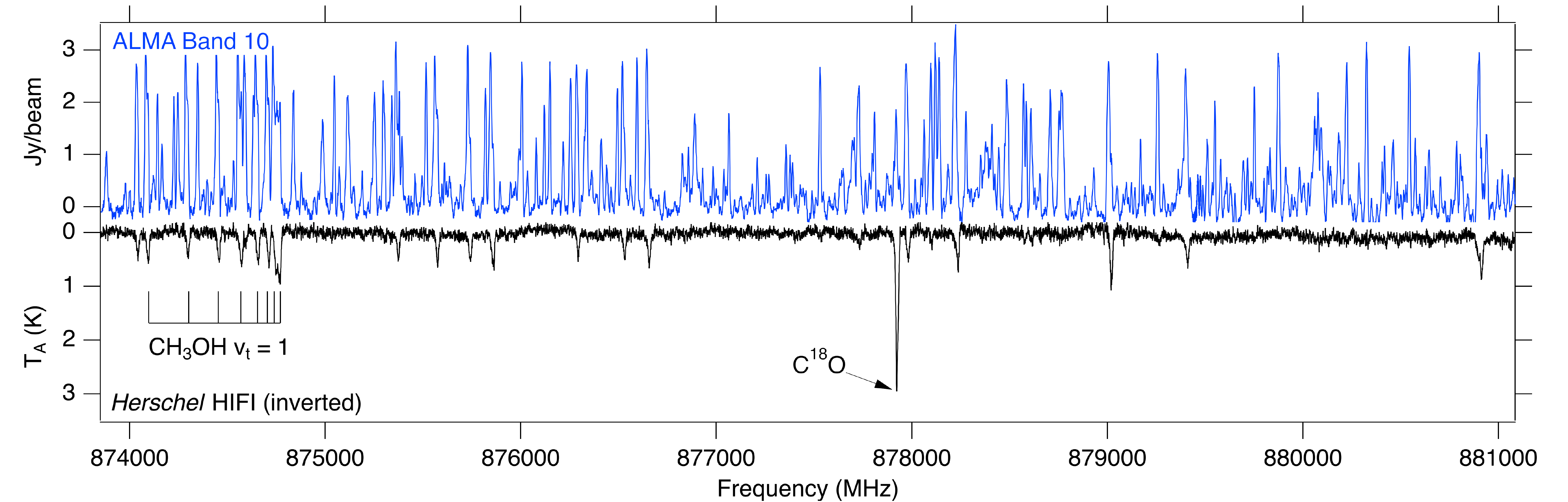} 
  \caption{ALMA Band 10 spectrum of the high-mass star-forming region
    NGC 6334I, compared with the spectrum over the same frequency
    range obtained with {\it Herschel}. Note the huge number of lines
    detected with ALMA due to its much smaller beam. Around 20-70\% of
    ALMA lines are still unidentified, stressing the need for more
    laboratory spectroscopy. Figure based on \citet{McGuire18b}.}
\label{fig:ALMAlines}
\end{center}
\end{figure}

{\underline {\it Spectroscopy.}}  High-mass star-forming clouds like
Orion-KL and SgrB2(N) are well known as line-rich sources. {\it
  Herschel} and ALMA line surveys reveal the rich chemical composition
of warm gas, but making a full inventory is hampered by the sometimes
large fraction of unidentified lines.  For {\it Herschel}, about
5-12\% of the channels are unidentified \citep{Crockett14,Neill14}.
This fraction is much higher for ALMA data
(Fig.~\ref{fig:ALMAlines}). The low-mass source IRAS 16293-2422B has
very narrow lines which allows identification of prebiotic molecules
toward solar-mass protostars for the first time
\citep{Jorgensen16,Jorgensen18}. Interestingly, in the 350 GHz window
(ALMA Band 7), the fraction of U-lines is about 20\%, but this grows
to $\sim$70\% in very deep observations at lower frequencies (230 and
115 GHz, ALMA Band 6 and 3) \citep{Taquet18}. Most of the lines are
likely due to isotopologs and/or vibrationally excited states of known
complex organic molecules. For example, the abundant HNCO molecule
shows lines of HN$^{13}$CO, HNC$^{18}$O, HNC$^{17}$O, H$^{15}$NCO,
DNCO, DN$^{13}$CO, DNC$^{18}$O and many more combinations, for which
laboratory data were still lacking until recently. This, in turn,
prevents identification of new species, including amino acids. ALMA
spectra of AGB stars also contain a large fraction of U-lines
\citep{Cernicharo13} that may be due to different types of species
(e.g., metal-containing), especially if they are probing the dust
formation zone. Databases such as the Cologne Database for Molecular
Spectroscopy \citep{Endres16} and the JPL molecular database are
invaluable resources for identifying lines \citep[see review
  by][]{Widicus19}.

{\underline {\it Collisional rate coefficients.}} Considerable
progress continues to be made in this area through quantum chemical
calculations, thanks also to coordinated efforts maintaining a close
link with observers. State-to-state collisional rate coefficients with
H$_2$ for pure rotational transitions have been computed in the last
decade for H$_2$O, H$_2$CO, HCN, HNC, CN, CS, SO, SO$_2$, CH, CH$_2$,
HF, HCl, OH$^+$, NH$_2$D, HC$_3$N, CH$_3$CN, and CH$_3$OH, among
others. For ions and molecules with large dipole moments such as HF,
CH$^+$ and ArH$^+$, collisions with electrons are also
important. Driven by infrared spectra from {\it ISO} and {\it
  Spitzer}, and with {\it JWST} on the horizon, there is now also
attention to the calculation of rate coefficients for
vibration-rotation transitions.  For example, new data have been
provided for CO with H$_2$ and H \citep{Song15}. All of these
data are very time consuming to compute, sometimes taking nearly a
decade per species (e.g., H$_2$O). The data can be accessed through
the BASECOL database \citep{Dubernet13} and the LAMDA database
\citep{Schoier05,vanderTak07}.

{\underline{\it Chemical reactions.}} Rate coefficients for
gas-phase chemical reactions are derived from experiments and from
theory. The main databases are UMIST \citep{McElroy13} and KIDA
\citep{Wakelam15}, which continue to be updated with new results from
the chemical physics literature.
An interesting example is the associative detachment reaction, H$^-$ +
H $\to$ H$_2$ + e, which controls the amount of H$_2$ in the early
Universe and thus its cooling. This reaction was studied in the 1970s
but never again until this decade \citep[e.g.,][]{Kreckel10}; the new
values significantly reduce the uncertainty in the masses of the first
stars found in hydrodynamical simulations.

{\underline{\it Photodissociation.}} UV radiation is one of the main
destruction routes of molecules. \citet{Heays17} have provided an
update of the wavelength dependent photodissociation and
photoionization cross sections for many astrophysically important
molecules and atoms.  Rates are provided both for the interstellar
radiation field, for cool stars and the Sun, as well as for the cosmic
ray induced field (important inside dark cores and disks). The
photodissociation of CO and N$_2$ and their isotopologs continues to
be studied experimentally, leading to refinements in self-
and mutual shielding factors \citep{Visser09,Heays14}.  Taking the
wavelength dependence into account is particularly important for the
chemistry in protoplanetary disks around different types of stars.

{\underline{\it Other types of data.}} One example are Land\'e
factors, needed to measure magnetic fields in interstellar clouds
through Zeeman splitting of molecules. In star-forming regions,
methanol is particularly abundant and has strong lines, both thermally
excited and as masers. Calculation of the CH$_3$OH Land\'e factors by
\citet{Lankhaar18} allowed the magnetic field in the Cep A cloud to be
measured at $7.7\pm 1.0$ mG. Other examples can be found in
\S~\ref{sec:exoplanets}.

\section{Molecular physics: PAHs, fullerenes}
\label{sec:PAH}

Strong and broad emission features at mid-infrared wavelengths have
been seen since the 1970s throughout the ISM and are most plausibly
ascribed to a collection of Polycyclic Aromatic Hydrocarbons (PAHs) of
50--100 carbon atoms each \citep[see review by][]{Tielens08}. Since
these features can even be detected out to high redshift and are
excited by UV radiation, they can be widely used as diagnostics, for
example to measure star-formation rates in obscured regions. More
subtle variations in band shapes and relative strengths are seen in
local sources, such as PDRs versus disks or evolved stars
\citep{Peeters02,Peeters17}. To unlock the potential of PAHs as
diagnostics, a large range of data is needed, from 
spectra of individual PAHs \citep{Boersma14,Mackie15} to their
ionization and electron recombination/attachment rates
\citep[e.g.,][]{LePage01} and dehydrogenation rates
\citep[e.g.,][]{Bouwman19}.

Individual PAHs have not yet been identified through infrared
spectroscopy: only their ionization stage (positively or negatively
charged vs.\ neutral) and their global size and molecular structure
have been characterized. Small PAHs with a side group and dipole
moment also have strong radio spectra, however. The recent GBT
detection of one of the smallest PAHs, C$_6$H$_5$CN (benzonitrile), at
centimeter wavelengths in a cold dark cloud, TMC-1, therefore caused
considerable excitement \citep{McGuire18a}. This detection opens the
door for searches for other PAHs and also raises interesting questions
on its formation, which is most likely bottom-up starting from
C$_2$H$_2$, rather than top down. In addition to experiments, quantum
chemical studies can be useful to identify reaction paths and stable
structures starting from C$_2$H$_2$ \citep[e.g.][]{Peverati16}.

A subset of infrared bands between 5 and 22 $\mu$m can be uniquely
assigned to fullerenes, C$_{60}$ and C$_{70}$
\citep{Cami10}. Initially identified in a young planetary nebula, they
are now seen in a wide variety of sources. Interstellar C$_{60}^+$,
originally proposed by \citet{Foing94}, is now firmly detected at
optical wavelenths \citep{Cordiner19}. In fact, there is now good
observational evidence for the transformation of PAHs to fullerenes
under the influence of UV radiation, a process that is also understood
theoretically \citep{Berne12} and simulated experimentally
\citep{Zhen14}.

 \begin{figure}[t]
\begin{center}
\includegraphics[width=11cm]{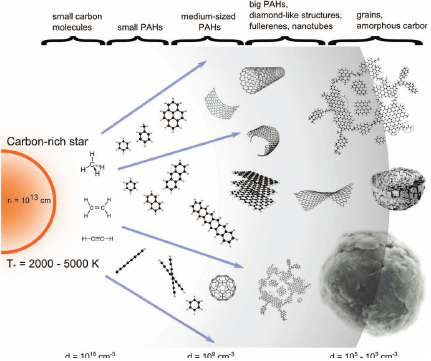} 
\vspace*{0.2 cm}
  \caption{Illustration of growth of carbonaceous molecules and solids in the envelopes of evolved stars. Figure from \citet{Contreras13} with permission.}
\label{fig:carbon}
\end{center}
\end{figure}

\newpage

\section{Condensed matter physics: solids, ices}

{\underline{\it Carbonaceous grains.}} When PAHs, fullerenes and other
carbon-containing species grow to large enough sizes (few hundred
atoms), they start to behave like solids rather than individual
molecules \citep[][Fig.~\ref{fig:carbon}]{Contreras13}. The structure
of interstellar carbonaceous material is amorphous but the details and
composition are under discussion, such as the fraction of aromatic
vs.\ aliphatic bonds: does the material resemble coal or hydrogenated
amorphous carbon? A related question is how and where they are
formed. It has been known for some time that the destruction rates of
interstellar grains are larger than their production rates in
envelopes of evolved stars. Thus, some (re-)formation of
grains must also take place in the low density cold ISM. Laboratory
experiments are now starting to provide some proof that this indeed
can happen \citep{Fulvio17}.

{\underline{\it Silicate grains.}} The same issue also holds for the
formation of silicate grains. SiO polymerization has been shown to
take place without barriers in experiments using the He-droplet
technique to confine SiO molecules. This then provides a route for
producing silicate grains in the cold ISM by accretion onto a kernel
\citep{Krasnokutski14}.

Most interstellar silicate grains are amorphous but a distinct set of
peaks have been observed in protoplanetary disks and evolved stars,
that can be ascribed to crystalline silicates such as forsterite
(Mg$_2$SiO$_4$).  Much laboratory work has been carried out to make
these identifications \citep[see review by][]{Henning10} and more work
is needed since some of the longer wavelength features are highly
sensitive to precise composition and/or temperature \citep{Sturm10}.

{\underline{\it Ices: freeze out and sublimation.}}  At low dust
temperatures of $\sim$10 K, the probability for atoms and molecules to
freeze out on dust grains is unity on every collision, thus forming an
icy layer on top of the silicate or carbonaceous core. Once on the
grain, new molecules can form, ranging from simple species like H$_2$O
to complex organic molecules (see below). When the grain heats up,
these molecules will sublimate back into the gas in a sequence
according to their binding energies. Thus, laboratory determinations
of binding energies of ices, either as pure or mixed ices, are
essential for understanding of observations of star- and
planet-forming regions \citep{Collings04}.

Surprisingly, some molecules produced on grain surfaces are observed
in the gas at temperatures well below that for thermal sublimation
\citep[e.g.,][]{Bacmann12}. How to get molecules off the grains at low
temperatures intact is a major puzzle which is being addressed by
experiments and theory. One option is photodesorption by UV radiation
although this process often also dissociates the molecule (except for
CO and N$_2$) resulting in desorbing fragments \citep[see review
  by][]{Oberg16}.  Other options are chemical desorption, in which
part of the energy liberated by formation of the chemical bond is used
to desorb the molecule \citep{Minissale16}, or impulsive spot heating
by cosmic rays \citep{Ivlev15}.

 \begin{figure}[t]
\begin{center}
\includegraphics[width=10cm]{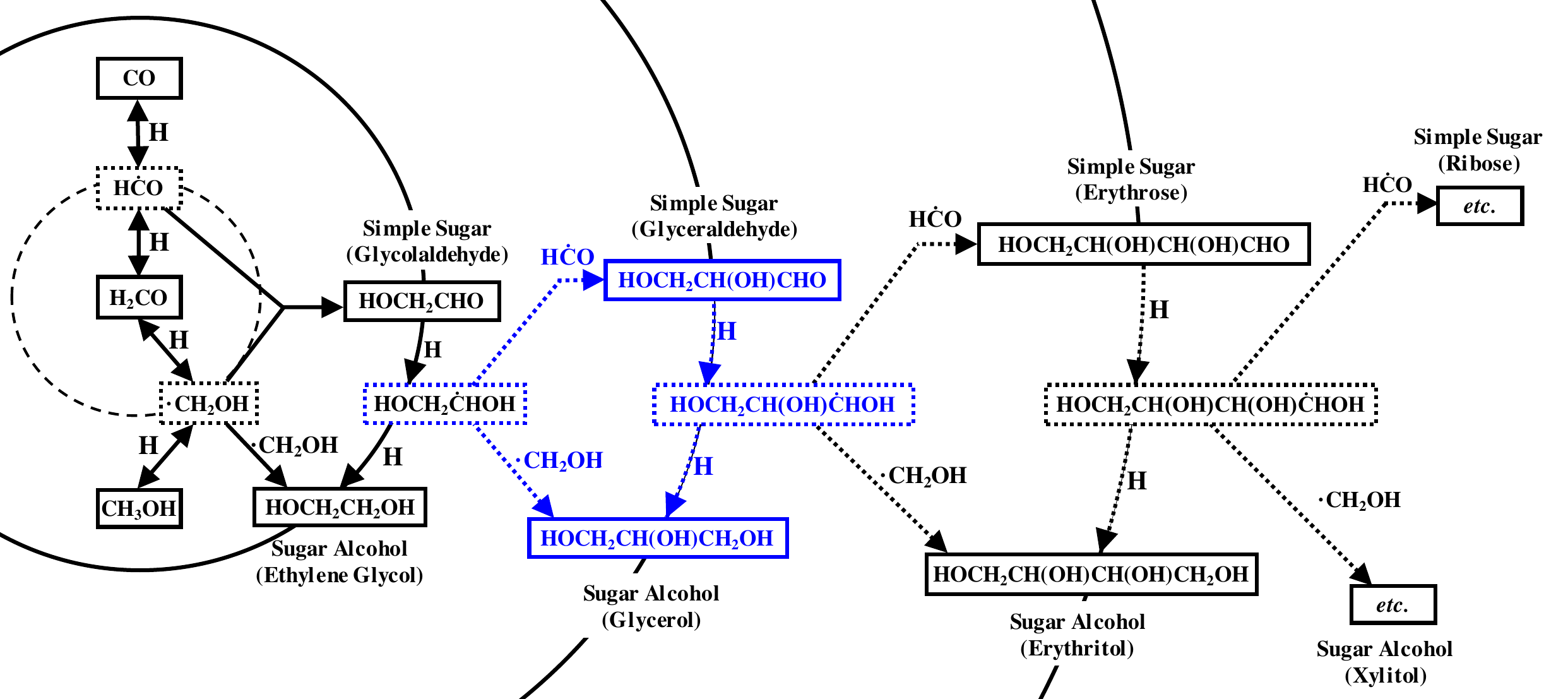} 
\vspace*{0.3 cm}
  \caption{Reaction scheme leading to complex sugars in ices at
    low temperatures (10-15 K), without the need for UV radiation or
    cosmic rays. Figure from \citet{Fedoseev17}.}
\label{fig:glycerol}
\end{center}
\end{figure}

{\underline {\it Ices: formation of complex organic molecules.}}  Many
complex molecules can be produced by reactions occuring on and in
ices.  Ultra-high vacuum solid-state laboratory experiments have
demonstrated that complex molecules like methanol and ethylene glycol
can indeed be formed at temperatures as low as 10 K without any energy
input, as long as some H atoms are available
\citep[e.g.,][]{Chuang16}. Starting from CO, the sequence may even go
as far as tri-carbon molecules like glycerol and real `sugars'
\citep{Fedoseev17} (Fig.~\ref{fig:glycerol}).  UV irradiation of ices
breaks bonds and produces radicals which become mobile upon heating,
leading to further molecular complexity \citep[e.g.,][]{Oberg16}. High
energy particle irradiation produces many of the same species, with
the difference that the strong CO and N$_2$ bonds can also be broken
for energies $>$11 eV.

The translation of these experiments into numbers that can be used in
astrochemical gas-grain models \citep[e.g.,][]{Garrod08}, such as
binding energies and diffusion and reaction barriers, is far from
simple, however.  Diffusion of at least one of the two reactants is
particularly important to make the reaction go, unless the two
reactants happen to be next to each other. If not, a prescription to
hop from site to site is needed, in which the crucial parameter is the
barrier $E_{\rm hop}$. Usually $E_{\rm hop}= c^{\rm st} \times E_{\rm
  bind}$ is assumed with c$^{st}$ varying between 0.3 and 0.7,
depending on species and surface site. The importance of tunneling at
low temperatures is still debated. A detailed overview of these issues
is presented in \citet{Cuppen17} and highlights the fact that some
fundamental chemical physics questions regarding surface chemistry are
still poorly understood.

\section{Planetary sciences: from dust to exoplanets}
\label{sec:exoplanets}

{\underline{\it From dust to pebbles and planetesimals.}}
Laboratory experiments using drop towers and other set-ups have
characterized and quantified the collisional growth of dust grains
from sub-$\mu$m size to pebbles and beyond over the past 25 years
\citep{Blum18}. Small silicate particles of various sizes are made to
collide with each other, with the outcome monitored as a function of
collision speed. Besides sticking, bouncing and fragmentation, other
processes such as erosion or mass transfer occur. When these lab
results are put into coagulation simulations, it is found that
$\mu$m-sized dust grains can indeed grow to mm- to cm-sized aggregates
before they encounter the bouncing barrier.  The effects of
ice-covered grains are still being debated and more experiments are
needed, ideally carried out in zero gravity. To grow beyond pebble
size, other processes such as gravitational collapse of dust
``pebbles" by the streaming instability is often invoked.

{\underline {\it Analyzing samples from space.}}
Extraterrestrial samples provide a wealth of information on the
formation processes in the early Solar System, including time scales
for specific rock formation events, the size distribution of
particles, and the level of compositional mixing on small and large
scales. Studies can range from in-situ analyses, such as done for
comet 67P/C-G with the {\it Rosetta} mission, to detailed
characterization of interplanetary dust particles collected with
airplanes and meteorites that have fallen on Earth. Sample return
missions that bring back rocks on Earth, ranging from the Lunar
Missions to the {\it Stardust} mission to comet Wild-2 and culminating
now with the {\it Hayubusa2} and {\it Osiris-Rex} missions to
asteroids, provide unique in-depth insight. This is because these
samples can be analyzed with increasingly sophisticated experiments
that can image and study the composition of the material on
nanometer scales: big machines such as NanoSIMS cannot be flown in
space. This of course also requires laboratories that are committed to
the long-duration ($>$50 yr!)  curation of the extraterrestrial
samples.

{\underline {\it (Exo)planetary atmosphere studies.}}
Studies of the atmospheres of our Solar System planets and their moons
(e.g., Titan) have long recognized the importance of various molecular
processes such as high-temperature (photo-)chemistry and pressure
broadening in modeling the data collected by various missions. These
needs are now (re-)emphasized for exoplanetary atmosphere studies,
which cover an even wider range of physical conditions than found in
our own Solar System \citep{Moses14,Fortney19}. This includes (i)
High-resolution molecular line lists; (ii) Pressure broadening; (iii)
Collision-induced absorption, dimers; (iv) Haze and condensate
formation and associated opacities; (v) Chemical reaction rates; and
(vi) Photodissociation cross sections at high temperatures.

To illustrate a few examples, consider the line lists. The ExoMol
program uses computational chemistry (quantum calculation of potential
surfaces with associated energy levels and transition probabilities)
to provide high accuracy line lists for an increasing list of
molecules \citep{Tennyson16}. Each molecule has billions of lines and
may take a few years work of an experienced researcher to compute. The
larger the molecule, the more time-consuming the calculations
become. Alternative, faster but less accurate methods are now being
developed (Sousa-Silva, priv.\ comm.). The HITRAN database collecting
experimental data remains invaluable as well \citep{Gordon17}.

 \begin{figure}[t]
\begin{center}
\includegraphics[width=8cm]{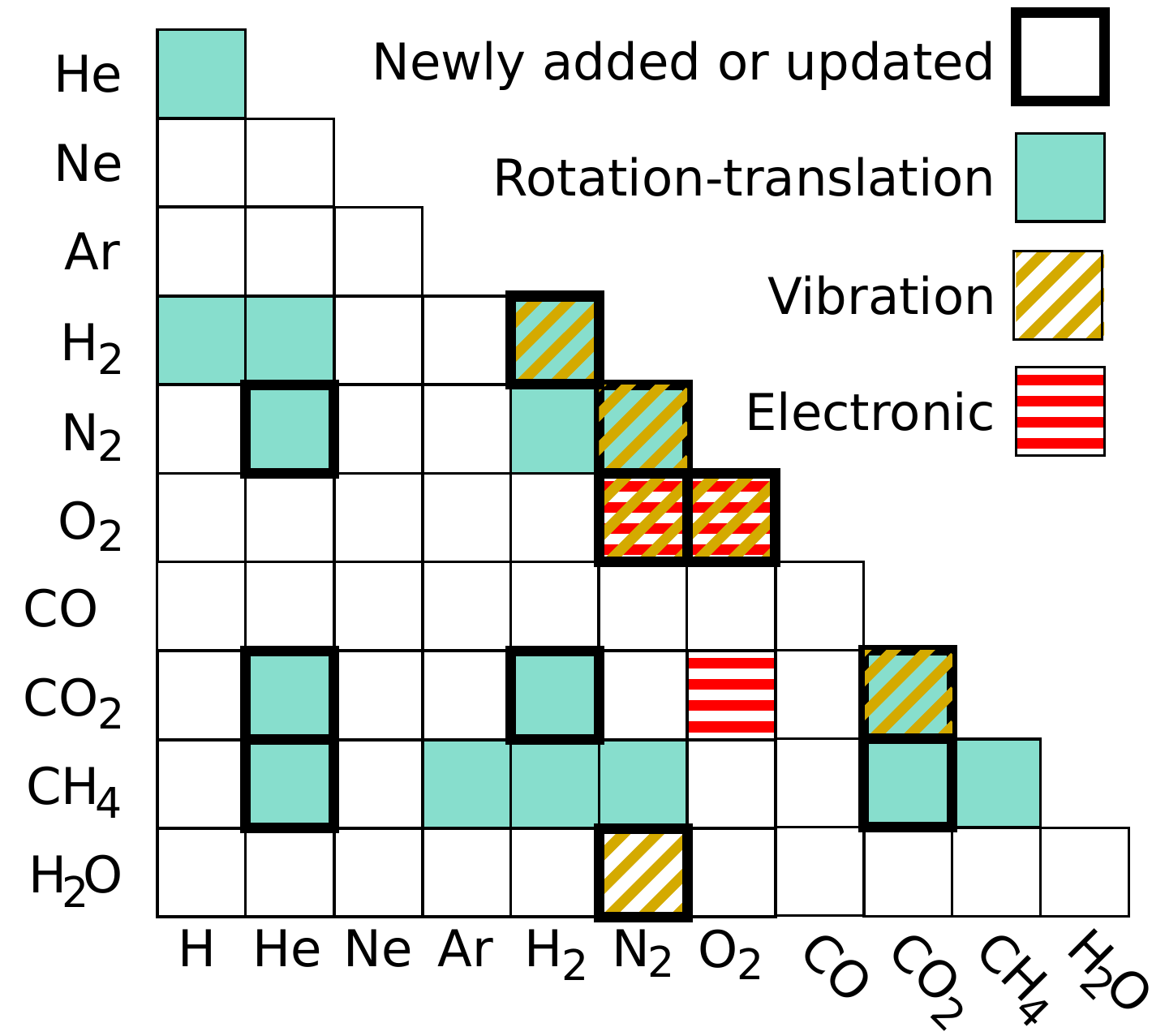} 
\vspace*{0.0 cm}
  \caption{Overview of systems for which collisional induced
    absorption data are available from calculations. Figure adapted
    from \citet{Karman19}.}
\label{fig:cai}
\end{center}
\end{figure}

Collision-induced absorption is a process that contributes
significantly to the total absorption of radiation in dense planetary
atmospheres. Fig.~\ref{fig:cai} provides an overview of systems for
which such data are now available \citet{Karman19}. It also shows that
data for a large number of important species are still missing.

The formation of clouds and hazes significantly affects the detection
of molecular bands in exoplanet and brown dwarf atmosphere
spectra, making them look `flat' or `grey'
\citep[e.g.,][]{Kreidberg14}. Exactly how these particles form and
under which conditions is still poorly understood, and their opacities
are largely unknown \citep{Marley15}. This will be a major
challenge for future laboratory experiments.

\section{Outlook}

The Universe continues to provide a unique laboratory in which to
study basic physical and chemical processes, raising fundamental
questions and new insight into these fields, off the well-beaten
track. Conversely, the deluge of data from new observational
facilities across a wide range of wavelengths continues to require
accurate atomic, molecular, particle and solid-state data under a huge
range of physical conditions. Continued interdisciplinary
collaborations between these fields are important to reap the full
scientific benefit from these billion $\$$, Euro or Yen facilities. A
comparatively small investment can go a long way, but prioritization
of data requirements are needed. Also, astronomers have to come to
terms with the fact that a significant fraction funding for such
experiments have to be provided by astronomy rather than physics or
chemistry. Joint approaches to funding agencies may be needed.

In the next 25+ years, with JWST and XRISM being launched in
2021--2022, Ariel in 2028 and Extremely Large Telescopes becoming
operational in the mid-2020s, there is a growing data need.  The IAU,
and in particular Commission B5 on Laboratory Astrophysics and
Commission H2 on Astrochemistry \footnote{\tt
  www.iau.org/science/scientific$_{-}$bodies/commissions}, can play a
role in bringing astronomers, chemists and physicists together to make
all these experiments and calculations happen!









\end{document}